\documentclass[prb,twocolumn,showpacs,amsmath,amssymb,superscriptaddress]{revtex4}
\usepackage{graphicx}
\begin{document}
\preprint{}
\title {Effect of pressure on the magnetostructural
transition in SrFe$_{2}$As$_{2}$}
\author{M. Kumar}\email{manoj.kumar@cpfs.mpg.de}
\affiliation{Max Planck Institute for Chemical Physics of Solids,
N\"{o}thnitzer Str. 40, 01187 Dresden, Germany}
\author{M. Nicklas}\email{nicklas@cpfs.mpg.de}
\affiliation{Max Planck Institute for Chemical Physics of Solids,
N\"{o}thnitzer Str. 40, 01187 Dresden, Germany}
\author{A. Jesche}\affiliation{Max Planck Institute for Chemical Physics of Solids,
N\"{o}thnitzer Str. 40, 01187 Dresden, Germany}
\author{N. Caroca-Canales}\affiliation{Max Planck Institute for Chemical Physics of Solids,
N\"{o}thnitzer Str. 40, 01187 Dresden, Germany}
\author{M. Schmitt}
\affiliation{Max Planck Institute for Chemical Physics of Solids,
N\"{o}thnitzer Str. 40, 01187 Dresden, Germany}
\author{M. Hanfland}
\affiliation{ESRF, BP 220, 38043 Grenoble Cedex 9, France}
\author{D. Kasinathan}
\affiliation{Max Planck Institute for Chemical Physics of Solids,
N\"{o}thnitzer Str. 40, 01187 Dresden, Germany}
\author{U. Schwarz}\affiliation{Max Planck Institute for Chemical Physics of Solids,
N\"{o}thnitzer Str. 40, 01187 Dresden, Germany}
\author{H. Rosner}\affiliation{Max Planck Institute for Chemical Physics of Solids,
N\"{o}thnitzer Str. 40, 01187 Dresden, Germany}
\author{C. Geibel}
\affiliation{Max Planck Institute for Chemical Physics of Solids,
N\"{o}thnitzer Str. 40, 01187 Dresden, Germany}



\date{\today}
\begin{abstract}
We present a systematic pressure study of poly- and single
crystalline SrFe$_2$As$_2$ by electrical resistivity and X-ray
diffraction measurements. SrFe$_2$As$_2$ exhibits a structural phase
transition from a tetragonal to an orthorhombic phase at $T_0=205$
K. The structural phase transition is intimately linked to a
spin-density-wave transition taking place at the same temperature.
Our pressure experiments show that $T_0$ shifts to lower
temperatures with increasing pressure. We can estimate a critical
pressure of 4 to 5~GPa for the suppression of $T_0$ to zero
temperature. At pressures above 2.5~GPa the resistivity decreases
significantly below $T_x\approx40$~K hinting at the emergence of
superconductivity but no zero-resistance state is observed up to 3
GPa.

\end{abstract}


\maketitle


\section{Introduction}
The recent discovery of superconductivity and the subsequent raising
of the superconducting transition temperature, $T_{c}$, in iron
arsenides has attracted tremendous interest in the scientific
community. Superconductivity was first reported in \textit{R}FeOAs
compounds (\textit{R} = La -- Gd)\cite{kamihara, xchen, cruz,
klauss,
  fratini} and later on in the \textit{A}Fe$_2$As$_2$ series of
compounds (\textit{A} = Ca, Sr, Eu,
Ba).\cite{ni,krellner,rotter1,jeevan} The structures of both
families of compounds present almost identical FeAs-layers which are
responsible for the peculiar behavior of these systems. The undoped
compounds show a tetragonal to orthorhombic structural phase
transition associated with magnetic ordering giving rise to a
spin-density-wave instability between $150-200$~K,\cite{ni,krellner,
sasmal, gchen1} which can be suppressed by chemical substitution or
application of pressure.\cite{sasmal,rotter1, torikachvili}
Doping-induced superconductivity was observed in
fluorine-doped/O-deplated members of \textit{R}FeOAs with $T_{c}$
values approaching up to $50$~K.\cite{kamihara, xchen, fratini} The
superconducting transition temperature reaches up to 38 K in
\textit{A}Fe$_2$As$_2$ (\textit{A} = Sr, Ba) on partial replacement
of \textit{A} with K or Cs.\cite{sasmal,rotter1,gchen2} However,
chemical substitution in contrast to the application of pressure
changes not only the unit-cell volume but also the electronic
structure considerably for example by adding or removing charge
carriers from the conduction band.

In CaFe$_2$As$_2$ pressure studies reported a very fast suppression
of the transition temperature $T_0$ related to the magnetic ordering
and lattice distortion and its disappearance at around
0.4~GPa.\cite{torikachvili,park} Simultaneously, superconductivity
appears with a maximum $T_c \approx 12$~K. Further on an anomaly
appears above 0.5~GPa at around 100~K and shifts strongly to higher
temperatures with increasing $p$. Neutron scattering experiments
evidenced this anomaly to correspond to a structural transition
towards a collapsed, tetragonal structure, with a reduced $c/a$
ratio.\cite{Kreyssig} In contrast measurements on BaFe$_2$As$_2$
revealed a much weaker pressure effect, with $T_0$ decreasing by
only 15~K at 2~GPa, and no superconductivity until this
pressure.\cite{torikachvili2} A much weaker pressure effect for the
larger earth alkaline metals Sr and Ba was confirmed by Alireza
\textit{et al.}\cite{alireza} who reported the onset of
superconductivity at 2.8~GPa and 2.5~GPa with maximum $T_c = 27$ K
and 29 K for SrFe$_2$As$_2$ and BaFe$_2$As$_2$, respectively.
However, one has to note that all these results were obtained on
single crystals grown from Sn flux. Especially in the case of
BaFe$_2$As$_2$ this technique is known to lead to a significant
incorporation of Sn into the single crystals, which strongly affects
the physical properties, resulting e.g. in a strong reduction of
$T_0$. On the other hand the huge difference between the pressure
effect on CaFe$_2$As$_2$ compared to that on SrFe$_2$As$_2$ and
BaFe$_2$As$_2$, as well as the absence of a collapsed phase in doped
\textit{A}Fe$_2$As$_2$ (and doped \textit{R}FeAsO) suggests that
this collapsed phase might be unique to CaFe$_2$As$_2$, being
related to the comparatively small size of Ca, and not a general
feature of the \textit{A}Fe$_2$As$_2$ materials.  The absence of a
collapsed phase in superconducting, doped \textit{A}Fe$_2$As$_2$
already proves that this collapsed phase is neither a prerequisite
for the disappearance of magnetic order, nor for the onset of
superconductivity.

X-ray studies on SrFe$_2$As$_2$ at room temperature and down to
210~K evidenced an undistorted tetragonal ThCr$_2$Si$_2$ type
structure while at 205~K and below the diffraction diagrams can be
well described by an orthorhombic unit cell in accordance with the
proposed structure for BaFe$_2$As$_2$. Between 210~K to 205~K, the
high-temperature undistorted tetragonal phase disappears abruptly
but a small amount of the orthorhombic phase coexists as expected
for a first-order phase-transition. Resistivity, $\rho(T)$,
susceptibility, $\chi(T)$, and specific heat, $C(T)$, measurements
show anomalies around 205 K.\cite{anton} Our results on poly- and
single crystalline SrFe$_2$As$_2$ from high pressure electrical
resistivity and X-ray diffraction experiments reveal a suppression
of $T_0$ with increasing pressure and hint to the possible existence
of a magnetic instability in the pressure range from 4 to 5~GPa.
Resistivity data suggests the emergence of a superconducting phase
at $p>2.5$~GPa.

\begin{figure}[t]
\centering
\includegraphics[angle=0,width=8cm,clip]{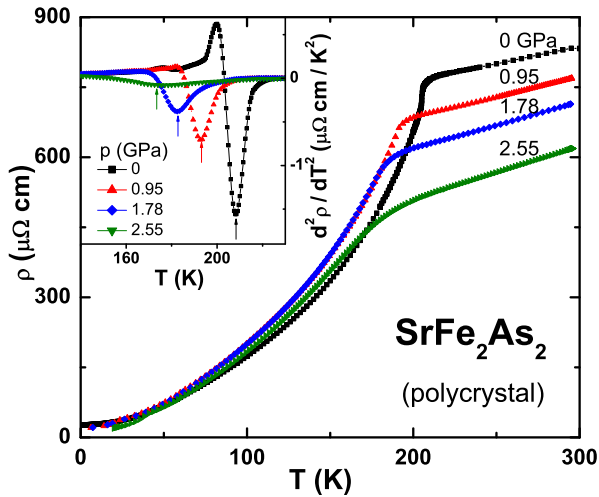}
\caption{\label{fig1} (Color online) Electrical resistivity {\it
vs.} temperature for different pressures for polycrystalline
SrFe$_2$As$_2$. Inset: ${\rm d^2}\rho(T)/{\rm d}T^2$ for the same
pressures, arrows indicate $T_0$.}
\end{figure}

\section{METHODS}

The polycrystalline SrFe$_2$As$_2$ samples were synthesized by
heating a $1:2:2$ mixture of Sr, Fe, and As in an Al$_2$O$_3$
crucible, sealing it under inert atmosphere inside an evacuated
quartz tube and subsequent heating.\cite{krellner} Single crystals
were obtained using the Bridgman method (cf.
Ref.~\onlinecite{jeevan}). Both types of samples crystallize in the
undistorted tetragonal ThCr$_2$Si$_2$ structure. Measurements of the
electrical resistance were carried out using a standard four-probe
technique.  Magnetic field was applied perpendicular to the current.
The experiments on single crystals were made with current flowing in
the $(a,b)$-plane and magnetic field along the $c$-direction.
Temperatures down to 1.8 K and magnetic fields up to 14~T were
generated using a flow cryostat and a physical property measurement
system (Quantum Design). Pressures up to 3~GPa have been achieved in
a double-layer piston-cylinder-type pressure cell with silicone oil
as pressure transmitting medium. The superconducting transition of
Pb which served as a pressure gauge, remained sharp at all
pressures, indicating a pressure gradient less than $1-2$~\% of the
applied pressure. The pressure change on cooling from room
temperature to 1.8~K was less than 0.1~GPa. X-ray diffraction under
pressure was conducted on ground crystalline powders of the
compound. The samples were loaded into the gasket of a
membrane-driven diamond anvil cell with a culet size of 0.6~mm. In
order to realize hydrostatic conditions, helium was used as a
pressure transmitting medium. A helium gas-flow cryostat enabled
thermostated low-temperature measurements. Sm-doped SrB$_4$O$_7$ was
used as a temperature-insensitive pressure calibrant.\cite{T1,T2}
Diffraction data were collected on ID9A at the ESRF, Grenoble, using
a wavelength of 41.34~pm. During exposure, samples were oscillated
by six degrees in order to enhance powder statistics. Finally, the
recorded 2D diffraction patterns were integrated by means of the
computer program Fit2D.\cite{fit2d} To study the pressure dependence
of the magnetic instability, band structure calculations have been
carried out within the local (spin) density approximation (L(S)DA)
fixing the As $z$ parameter to the experimental ambient pressure
values. We applied the full-potential local-orbital code
FPLO\cite{koepernik99} (V 7.00-28) with the Perdew-Wang exchange
correlation potential \cite{PW92} and a well-converged $k$-mesh of
$24^3$ points for the Brillouin zone.

\section{RESULTS AND DISCUSSION}

\begin{figure}[t]
\includegraphics[width=8cm,angle=0]{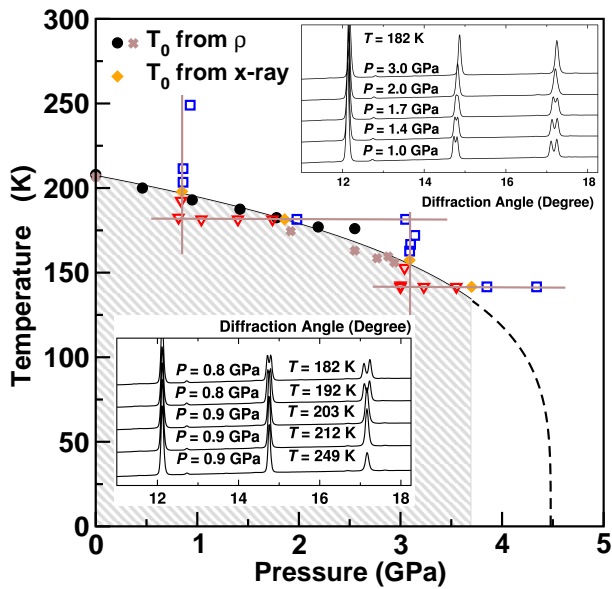}
\caption{\label{phasediagram} (Color online) Temperature-pressure
phase diagram of SrFe$_2$As$_2$ with $T_0$ values from the
evaluation of resistivity data (poly-crystal - filled circles,
single crystal -  stars) and X-ray measurements (filled diamonds).
The data points for the X-ray measurements are shown by open squares
and triangles for the tetragonal and the orthorhombic phase,
respectively. The approximately isobaric and isothermic runs are
guided by horizontal and vertical lines, exemplary X-ray patterns
are presented in the upper and lower inset. The splitting of the
reflections at about 15  and 17 degrees indicates the structural
transition. The measured  region for the orthorhombic (magnetic)
phase is shaded in gray, the  dashed line is an extrapolation of the
phase boundary down to zero  temperature.}
\end{figure}

Figure~\ref{fig1} shows the electrical resistivity of
polycrystalline SrFe$_2$As$_2$ as a function of temperature for
selected pressures. At ambient pressure $\rho(T)$ is weakly
decreasing between 300~K and 205~K. At around 205~K $\rho(T)$ shows
a sharp drop at $T_0$ followed by a further strong decrease to low
temperatures, in quite good agreement with the previously reported
$T_0=198$ K (Ref.~\onlinecite{yan}). On increasing pressure the
feature at $T_0$ is becoming broader but is still well defined up to
the highest pressure of our electrical resistivity experiment. There
is no qualitatively different behavior between the polycrystalline
and the single crystalline material. At ambient pressure the
polycrystals have a residual resistivity ratio of
$RR_{\rm1.8K}=\rho_{\rm300K}/\rho_{\rm1.8K}=32$ and a room
temperature resistivity $\rho_{\rm300K}=0.83~{\rm m\Omega cm}$
compared with $RR_{\rm1.8K}=25$ and $\rho_{\rm300K}=1.09~{\rm
m\Omega cm}$ for the single crystals. However, the kink at $T_0$
remains sharper for the single crystal at high pressure. It is also
worth mentioning that magnetic field of $B=9$~T has no effect on
$T_0$ in the single crystalline sample at our highest pressure of
2.94~GPa for $B\parallel c$, similar to the behavior reported at
ambient pressure.\cite{yan} $T_0$ as defined by the minimum in the
second derivative ${\rm d^2}\rho(T)/{\rm d}T^2$ (cf. inset of
Fig.~\ref{fig1}) shifts towards lower temperatures on application of
pressure. The phase diagram in Fig.~\ref{phasediagram} summarizes
the results. Within the error-bars there is no difference between
the poly- and single crystalline samples. Initially $T_0(p)$
decreases with a slope of ${\rm d}T_0/{\rm d}p\mid_{p=0}\approx
-13$~K/GPa with increasing pressure. ${\rm d}T_0/{\rm d}p\mid_{p=0}$
of SrFe$_2$As$_2$ is in fairly good agreement with ${\rm d}T_0/{\rm
d}p\mid_{p=0}\approx -10.4$~K/GPa reported for BaFe$_2$As$_2$
\cite{torikachvili2} but almost one order of magnitude smaller than
for CaFe$_2$As$_2$.\cite{torikachvili}

\begin{figure}[b]
\includegraphics[width=8cm,angle=0]{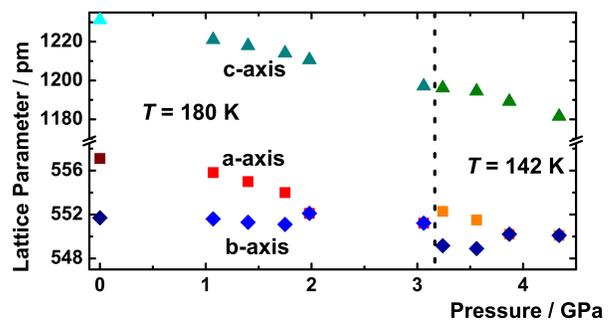}
\caption{\label{fig3} (Color online) Lattice parameters as refined
from X-ray diffraction data collected at different temperatures and
pressures. Coincidence of the parameters $a$ and $b$ indicate the
tetragonal phase.}
\end{figure}

In the \textit{A}Fe$_2$As$_2$ compounds, $T_0$ at ambient pressure
corresponds to both the structural transition and to the onset of
antiferromagnetic ordering, which are intimately linked
together.\cite{krellner,anton} In contrast, it is presently
suggested that in the \textit{R}FeAsO compounds,\cite{klauss} the
antiferromagnetic ordering occurs at approximately $10 - 20$~K below
the structural ordering. There both transitions are marked by an
anomaly in the resistivity.\cite{klauss} A careful analysis of our
resistivity data do not reveal any evidence for a splitting of the
anomaly at $T_0$ under pressure, neither in $\rho(T)$ nor in its
first or second derivative. Thus our resistivity data evidence that
both transitions remain linked together under pressure. In order to
get additional information on the effect of pressure on this
transition, we carried out temperature-dependent high pressure X-ray
diffraction studies for pressures up to 4.4~GPa and temperatures
down to 140 K. We performed two approximately isothermal and
isobaric runs, respectively. Exemplary X-ray patterns are presented
in the insets of Fig.~\ref{phasediagram} and the lattice parameters
of the isothermal measurements in Fig.~\ref{fig3}. For each run, we
could observe a clear phase transition from the tetragonal to the
orthorhombic phase with decreasing temperature or a suppression of
the distortion with increasing pressure. The phase boundary has been
determined as the midpoint (filled diamonds in
Fig.~\ref{phasediagram}) between the boundary points of the two
phases.

\begin{figure}[t]
\centering
\includegraphics[angle=0,width=8cm]{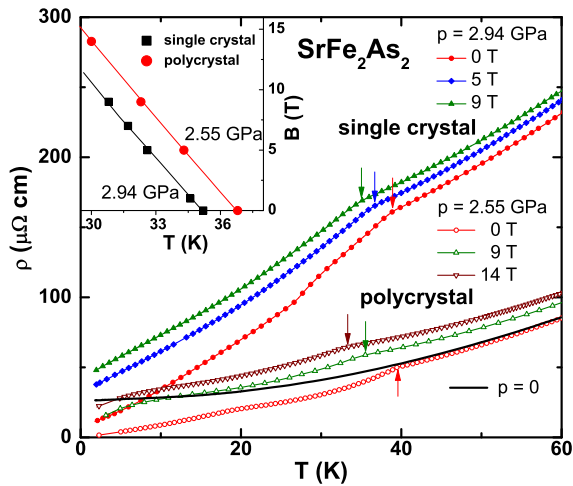}
\caption{\label{fig4} (Color online) Electrical resistivity of
SrFe$_2$As$_2$ at $p=2.55$~GPa (polycrystal, open symbols) and at
$p=2.94$~GPa (single crystal, solid symbols) in different applied
magnetic fields. The solid line represents the ambient pressure data
for the polycrystalline sample as a reference. The arrows indicate
$T_x$. Inset: $B-T$ diagram compiled from the resistivity data.}
\end{figure}

At 2.55~GPa a sharp decrease in resistivity shows up below 40~K in
the data of the polycrystalline sample. In the single crystal a
similar reduction is observed at a slightly higher pressure
($p=2.88$~GPa) and at somewhat lower temperature. However, we do not
observe zero-resistance at any pressure investigated in this study.
The transition temperature $T_x$ is in the same range where electron
doped SrFe$_2$As$_2$ becomes superconducting giving a hint at a
superconducting origin of the reduced resistance below
$T_x$.\cite{alj} To further elucidate the nature of the transition
observed in the electrical resistivity we applied a magnetic field.
The results at $p=2.55$~GPa for the polycrystalline and at
$p=2.94$~GPa for the single crystalline sample are presented in
Fig.~\ref{fig4}. With increasing magnetic field the reduction of the
resistivity is getting smaller and the $T_x$ shifts in the whole
accessible magnetic field range towards lower temperatures. Up to
14~T $B_x(T)=B(T=T_x)$ depends linearly on the magnetic field and
the initial slope ${\rm d}B_x(T)/{\rm d}T\mid_{T=T_x}=-2.05 {\rm~
T/K}$, the same for both samples and pressures, is typical for the
superconducting upper critical field, $B_{c2}(T)$, in the
iron-arsenide compounds.\cite{torikachvili} From these indications
we speculate that the observed drop in the electrical resistivity
indicates the emergence of a superconducting phase in
SrFe$_2$As$_2$.

Comparing the phase boundary constructed from the X-ray diffraction
and the electrical resistivity data for both poly- and single
crystalline samples, we find excellent agreement between all
measurements. Thus, we clearly observe an orthorhombic phase for
pressures below 3.8~GPa and $T<140$~K. This is most likely in
contrast to the data obtained by Alireza {\it et al.},\cite{alireza}
where superconductivity with $T_c\approx27$ K was observed for
SrFe$_2$As$_2$ at 2.8~GPa. At ambient pressure, the occurrence of
the orthorhombic phase is intimately linked to antiferromagnetism.
According to our electronic structure calculations, this intimate
connection between the antiferromagnetic order and the orthorhombic
distortion is preserved under pressure. Simulating hydrostatic
pressure in our calculations, we find that the magnetic instability
disappears at about 10 percent volume reduction, corresponding to a
critical pressure of slightly more than 10~GPa. This value should be
considered as a rough upper estimate, since it suffers from the
known LDA overestimate of magnetism in this class of compounds.  In
contrast to CaFe$_2$As$_2$, where our calculations indicate the
tetragonal collapsed phase similar to Ref.~\onlinecite{Kreyssig}, we
find no such transition for $A$Fe$_2$As$_2$ ($A$ = Sr, Ba, Eu).
This suggests that the $c/a$ collapse of the tetragonal phase is a
rather special feature of the CaFe$_2$As$_2$ system without general
relevance for the phase diagram of the $A$Fe$_2$As$_2$ compound
family. A more precise study, including the pressure dependence of
the magnetic transition upon doping, will be the subject of future
investigation.\cite{kasinathan08} Our preliminary results indicate a
considerable influence of doping and impurities on the critical
pressure. This result, although preliminary, may offer an
explanation for the observed differences of transition pressures in
different samples.\cite{torikachvili, park, alireza}

Results from specific heat, magnetic susceptibility and
resistivity,\cite{krellner} as well as X-ray, neutron diffraction,
muon-spin relaxation, and M\"{o}ssbauer experiments indicate a
first-order nature of the transition at $T_0$.\cite{anton} However,
Tegel {\it et al.}\cite{tegel} conclude a second-order type of the
transition from their temperature-dependent X-ray powder diffraction
and M\"{o}ssbauer spectroscopy. To get a further insight in the
nature of the phase transition, we analyzed the slope of our
$T_0(p)$ data at $p=0$ utilizing the Clausius-Clapeyron equation
${\rm d}T/{\rm d}p=T\Delta V/\Delta H$ applicable for a first-order
phase-transition. With the initial slope ${\rm d}T/{\rm
d}p=-13$~K/GPa and the latent heat at the transition $\Delta H
\approx200$~J/mol,\cite{anton} we obtain a volume change for the
orthorhombic unit cell $\Delta V=-0.8\times10^{-4}$~nm$^3$, which
has the same sign and the same order of magnitude as the
experimental result $\Delta V=-0.3\times10^{-4}$ nm$^3$.\cite{tegel}
An analysis for a second-order transition leads to a discrepancy of
at least one order of magnitude between calculated and observed
specific heat anomaly. Therefore this comparison supports the
first-order nature of the transition. \newline

\section{SUMMARY AND CONCLUSION}

In summary, we have determined the effect of pressure on the
structural and magnetic transition in SrFe$_2$As$_2$ using
electrical resistivity and X-ray diffraction measurements. We
observe a weak decrease of $T_0$ with increasing pressure with an
initial slope of $-13$~K/GPa and a bending towards lower
temperatures at higher pressures. Extrapolating these data and
assuming a continuous suppression of $T_0$ down to $T=0$ would lead
to a critical pressure of the order of 4.5~GPa. However, the
suspected first-order nature of the phase transition at $T_0$  makes
a classical critical end-point at a finite temperature more likely.
We still observe a transition to the orthorhombic phase at 3.8~GPa
below 140 K. Nevertheless already at around 2.5~GPa we observed in
$\rho(T)$ a kink at 40 K leading to a stronger slope ${\rm
d}\rho(T)/{\rm d}T$ at lower temperatures. This is suggestive of
superconductivity emerging at the disappearance of the structural
and magnetic transition. This interpretation is supported by the
linear shift of this anomaly to lower temperatures with applied
magnetic field. These experimental observations are supported by
results of band structure calculations which also indicate the
antiferromagnetic order to become unstable upon volume reduction.
Thus, in contrast to the observation in the high temperature
superconductors based on cuprates, in the layered FeAs systems the
suppression of magnetism and the onset of superconductivity do not
need electron or hole doping, but can be achieved without doping by
tuning the electronic states with pressure. The suppression of
magnetism upon applying pressure is in accordance with and a further
hint for an itinerant character of the magnetism, since for
localized magnetism one usually expects an enhancement of the
ordering temperature with pressure.

{\it Note added.---} While revising our manuscript we got knowledge
of a paper presenting a resistivity study of SrFe$_2$As$_2$ in the
pressure range up to 2 GPa. \cite{torikachvili3} In this work a
similar decrease of $\rho(T)$ below $T_x$, which we observe only
above 2.5 GPa, is already reported at smaller pressure.

\section*{ACKNOWLEDGEMENTS}It is a pleasure to thank K. Meier for the support of the X-ray
diffraction measurements.

\end{document}